\documentstyle[12pt]{article}

\font\twelve=cmbx10 at 15pt
\font\ten=cmbx10 at 12pt

\def\cit#1{$^{[#1]}$}

\def\pr{Phys.\ Rev.\ }
\def\prl{Phys.\ Rev.\ Lett.\ }
\def\pl{Phys.\ Lett.\ }
\def\etal{{\it et al}.}

\newcommand{\eeq}{\end{equation}}

\newcommand{\ba}{\begin{array}}
\newcommand{\ea}{\end{array}}

\newcommand{\rf}[1]{(\ref{#1})}

\begin{document}

\begin{titlepage}

\begin{center}

\renewcommand{\thefootnote}{\fnsymbol{footnote}}

{\ten Centre de Physique Th\'eorique\footnote{
Unit\'e Propre de Recherche 7061} - CNRS - Luminy, Case 907}
{\ten F-13288 Marseille Cedex 9 - France }

\vspace{1,5 cm}

{\twelve POSITIVITY CONSTRAINTS \\
FOR SPIN-DEPENDENT PARTON DISTRIBUTIONS}

\vspace{0.3 cm}

\setcounter{footnote}{0}
\renewcommand{\thefootnote}{\arabic{footnote}}

{\bf Jacques SOFFER}

\end{center}

\vspace{1,5 cm}

\setlength{\baselineskip}{24pt}

\centerline{\bf Abstract}

We derive new positivity constraints on the spin-dependent structure
functions of the nucleon. These model independent results reduce
conside\-rably their domain of allowed values, in particular for the
chiral-odd parton distribution $h_1 (x)$.

\setlength{\baselineskip}{15pt}

\vspace{1,5 cm}



\bigskip

\noindent Number of figures : 3

\bigskip

\noindent September 1994

\noindent CPT-94/P.3059

\bigskip

\noindent anonymous ftp or gopher: cpt.univ-mrs.fr

\end{titlepage}

\setlength{\baselineskip}{24pt}

The nucleon parton distributions are important physical quantities
which contain crucial informations about the fundamental properties
of the nucleon structure. A precise knowledge of the parton
distributions is also needed if one wants to explore hard scattering
processes at future hadron colliders. For many years,
spin-independent parton distributions have been accurately measured
in a large number of experiments, in particular deep inelastic
scattering, and they are now known in a wide kinematic range. The
experimental program going on at HERA will further increase this
kinematic domain with smaller $x$ and larger $Q^2$. The situation is
rather different for spin-dependent parton distributions whose
experimental determination has been improved only recently with some
new measurements\cit{1} of $g_1^p (x)$ and $g_1^n (x)$ both at CERN
and SLAC by means of proton and neutron polarized deep inelastic
scattering. These polarized structure functions provide us with some
insight into the quark (or antiquark) helicity distributions usually
called $\Delta q (x)$ (or $\Delta \overline q (x)$). But in addition
to the spin average quark distributions $q (x)$ and to these
helicity distributions $\Delta q (x)$, there is a third class of
quark distributions called transversity distributions and denoted
$h_1^q (x)$. These physical quantities which violate
chirality\cit{2,3,4} decouple from deep inelastic scattering but can
be measured in Drell-Yan processes with both beam and target
transversely polarized. So far there is no experimental data on these
distributions $h_1^q (x)$ (or $h_1^{\overline q} (x)$), but there are
some attempts to calculate them either in the framework of the MIT bag
model\cit{3} or by means of QCD sum rules\cit{5}.

The purpose of this letter is to use positivity to derive
model-independent constraints on $h_1^q (x)$ which will restrict
substantially the domain of allowed values\cit{6}. Similar constraints can
be obtained for higher-twist parton distributions, as we will see
below.

Let us consider quark-nucleon elastic scattering $q (h) + N (H) \to
q (h') + N (H')$ ($h, h'$ and $H, H'$ are the helicities of quark
and nucleon respectively) which is described in terms of five
$s$-channel helicity amplitude, denoted by $\langle h' H' | h
H \rangle$\cit{7}. In the forward direction, as a consequence of
helicity con\-servation, only three independent amplitudes are
non-vanishing, namely\break
$\varphi^s_1 = \langle + + | + + \rangle, \
\varphi^s_3 = \langle + - | + - \rangle$ and $\varphi^s_2 = \langle + - |
- + \rangle$, whose imaginary parts are simply related to total cross
sections by the optical theorem.

The forward quark-nucleon amplitude is a $4 \times 4$ matrix $M$
in the space where the basis states are $| + + \rangle,\ | + -
\rangle,\ | - + \rangle$ and $| - - \rangle$. Positivity requieres
that $a^+ M a \ge 0$ where ``$a$'' is any 4 component vector in this
space. This implies essentially three conditions\cit{8},
\begin{equation}\label{Im}
I m \varphi^s_1 |_{t = 0} \ge 0,\quad I m \varphi^s_3 |_{t = 0} \ge 0
\eeq
and
\begin{equation}\label{Imvarphi}
I m \varphi^s_3 |_{t = 0} \ge \left| I m \varphi^s_2 |_{t = 0} \right| .
\eeq
Now the  three quark distributions considered above
$q (x),\ \Delta q (x)$ (denoted $f_1 (x)$ and $g_1 (x)$ in
ref.~\cite{3}) and $h_1^q (x)$ are defined by the light-cone Fourier
transformation of bilinear quark operators between nucleon
states\cit{3}. In fact these quark distributions are related to the
corresponding $u$-channel quark-nucleon helicity amplitudes $\varphi^u_i$'s
which are simply obtained from the $\varphi^s_i$'s by quark line reversal and
we have
\begin{equation}\label{qx}
\ba{ll}
& q (x) = {1 \over 2} I m (\varphi^s_1 + \varphi^s_3) |_{t = 0} \, ,
\\[3mm]
& \Delta q (x) = {1 \over 2} I m (\varphi^s_3 - \varphi^s_1) |_{t = 0}
\, , \\[3mm]
& h_1^q (x) = {1 \over 2} I m \varphi^s_2 |_{t = 0} \, .
\ea
\eeq

Using eq.\rf{qx}, the constraints \rf{Im} and \rf{Imvarphi}  read in terms of
the parton distributions
\begin{equation}\label{inq}
q (x) \ge 0,\quad q (x) \ge | \Delta q (x) |
\eeq
and
\begin{equation}\label{Delta}
q (x) + \Delta q (x) \ge 2 | h_1^q (x) |.
\eeq
Whereas the first two constraints~\rf{inq} are familiar and quite
obvious, the third constraint~\rf{Delta}, which is much less trivial, was
ignored so far. We show in fig.1 the region allowed by eq.\rf{Delta}
which is half the region obtained by considering instead,
\begin{equation}\label{banal}
q (x) \ge | h_1^q (x) |,
\eeq
as proposed in ref.~\cite{3}.

Clearly the same constraint~\rf{Delta} holds for all quark flavor $q =
u, d, s$, etc... and for their corresponding antiquarks. Obviously
any theoretical model should satisfy these constraints. In a toy
model\cit{9} where the proton is composed of a quark and a scalar diquark
one obtains the equality in eq.\rf{Delta}\cit{10}. In the MIT bag model, let us
recall that these distributions read\cit{3}
\begin{equation}\label{bag}
q(x)=f^2(x)+g^2(x),\ \Delta q(x)=f^2(x)-1/3g^2(x)\ \hbox{and}\
h^q_1(x)=f^2(x)+1/3g^2(x)
\eeq
and they also saturate \rf{Delta}. In this case, we observe that
$h^q_1(x)\geq\Delta q(x)$ but this situation cannot be very general because
of eq.\rf{Delta}. As an example let us assume $h^q_1(x)=2\Delta q(x)$. Such a
relation
cannot hold for all $x$ and we see that eq.\rf{Delta}, in particular if
$\Delta q (x) > 0$, implies $q (x) \ge 3 \Delta q (x)$. This is certainly not
satisfied for all $x$ by the present determination of the $u$
quark helicity distribution, in particular for large $x$ where
$A_1^p (x)$ is large\cit{1}. The simplifying assumption $h_1^q (x) =
\Delta q (x)$, based on the non-relativistic quark model, which has been used
in some recent calculations\cit{11,12} is also not acceptable for all $x$
values if $\Delta q(x) < 0$ because of eq.\rf{Delta}. To illustrate the
practical use of eq.\rf{Delta}, let us take, as an example, the simple relation
\begin{equation}\label{bou}
\Delta u (x) = u (x) - d (x)
\eeq
proposed in ref.~\cite{13} and which is well supported by the
data\cit{1}. It is then possible to obtain the allowed range of
values for $h_1^u (x)$, namely
\begin{equation}\label{bour}
u(x)-{1 \over 2} d (x) \ge | h_1^u (x) |
\eeq
which is shown in fig.2, where ref.~\cite{13} was used to evaluate $u(x)$ and
$d (x)$. In this case, we see that for $x > 0.5$, both the results of
the MIT bag model\cit{3} and the QCD sum rule\cit{5} violate our
positivity bound, combined with low $Q^2$ data. A similar calculation can be
done for the $d$ quarks and the allowed region for $h_1^d(x)$ is shown in
fig.3.

We also want to remark that eq.\rf{Delta} puts a bound on the ''tensor charge''
$\delta q$ whose expression in terms of $h^q_1(x)$ and $h^{\bar q}_1(x)$ is
\begin{equation}
\int^{1}_{0} [h^q_1(x) -h^{\bar q}_1(x)]dx=\delta q\, .
\eeq
Since the sea quarks do not contribute to $\delta q$, as a consequence of
eq.\rf{Delta} one has
\begin{equation}
|\delta q|\leq {1\over 2}\int^{1}_{0} [q_{val}(x)+\Delta q_{val}(x)]dx\, .
\eeq
For $u$ quarks we get
\begin{equation}
|\delta u|\leq 1+{1\over 2}\int^{1}_{0}\Delta u_{val}(x)dx
\eeq
and for $d$ quarks
\begin{equation}
|\delta d|\leq {1\over 2}+{1\over 2}\int^{1}_{0}\Delta d_{val}(x)dx\, .
\eeq
By using the results of ref.\cit{13} one obtains
\begin{equation}
|\delta u|\leq {3\over 2}\ \hbox{and}\ |\delta d|\leq {1\over 3}\, .
\eeq

So far we have only considered the three twist-two quark (antiquark)
distributions, but the above results, and in particular eq.\rf{Delta},
are also valid for higher-twist distributions, which have been
identified in ref.~\cite{3}. So it is clear that we have the
following constraints for the twist-three distributions
\begin{equation}\label{ex}
e (x) + h_L (x) \ge 2 | g_T (x) |
\eeq
and for the twist-four distributions
\begin{equation}\label{fx}
f_4 (x) + g_3 (x) \ge 2 | h_3 (x) |,
\eeq
where we have used the notations of ref.~\cite{3}. There are
theoretical calculations based on the MIT bag model\cit{3,14} for
the twist-three distributions and we hope they satisfy the
constraint~\rf{ex}.

None of the above generalized distributions, which are associated to
quark-gluon dynamics, have been measured so far. As discussed in
ref.~\cite{3}, the most natural place to learn about them is
probably unpolarized and polari\-zed Drell-Yan and semi inclusive
processes. We hope much extensive studies both theoretical and
experimental will be undertaken in the future, where full
use of our new significant constraints~\rf{Delta}, \rf{ex} and \rf{fx} will be
made.

\section*{Acknowledgments}
The author would like to thank C. Bourrely, R.L. Jaffe, D. Sivers and T.T. Wu
for useful discussions at various stages of this work.

\newpage

\section*{Figure Captions}

\begin{itemize}

\item[Fig.1] The striped area represents the domain allowed by
positivity.

\item[Fig.2] The striped area represents the domain allowed for
$h_1^u (x)$ using eq.\rf{bour} and ref.~\cite{13}.

\item[Fig.3] The striped area represents the domain allowed for
$h_1^d (x)$ using eq.\rf{Delta} and ref.~\cite{13}.

\end{itemize}

\end{document}